\documentstyle[aps,prl,epsf]{revtex}
\begin{document}
\title{Zero-bias conductance anomaly in bilayer quantum Hall systems\footnote{To appear in the 
proceedings of the Fourth Physical Phenomena at High Magnetic Fields (PPHMF-IV) conference.}}
\author{Yogesh N. Joglekar and Allan H. MacDonald}
\address{Department of Physics,\\University of Texas at Austin, Austin, TX 78712, \\ 
Indiana University, Bloomington, IN 47405.\\}
\maketitle
\begin{abstract}
Bilayer quantum Hall system at total filling factor $\nu=1$ shows a rich variety of broken 
symmetry ground states because of the competition between the interlayer and intralayer Coulomb 
interactions. When the layers are sufficiently close, a bilayer system develops spontaneous 
interlayer phase-coherence that manifests itself through a spectacular enhancement of the 
zero-bias interlayer tunneling conductance. We present a theory of this tunneling conductance 
anomaly, and show that the zero-bias conductance is proportional to the square of the 
{\it quasiparticle} tunneling amplitude. 
\end{abstract}

%-----------------------------------------------------------------------------------------------%

\noindent{\it Introduction:} Bilayer systems consist of two 2D electron gases separated by a 
distance $d$ comparable to the 
typical distance between electrons within one layer, and have been intensely investigated over 
the past decade~\cite{demler}. Weakly disordered bilayer quantum Hall systems at total filling 
factor $\nu=1$ undergo a quantum phase transition from a compressible state to a $\nu=1$ quantum 
Hall state with spontaneous interlayer phase-coherence when the layers are sufficiently 
close~\cite{demler}. It is conventional to use a pseudospin to represent the layer index, where 
pseudospin ``up'' denotes a state in the top layer, and pseudospin ``down'' denotes a state in 
the bottom layer. In the pseudospin language, the phase-coherent state can be viewed as an 
easy-plane ferromagnet with its pseudospin polarization $\vec{M}$ along the $x$-axis, $M_x =1$. 
In analogy with ferromagnets, the low-lying excitations of a phase-coherent bilayer system are 
long-wavelength fluctuations of the ordered moment, {\it i.e.} pseudospin waves. These collective 
modes transfer charge from one layer to the other, and have a linear dispersion at long 
wavelengths when the interlayer tunneling amplitude $\Delta_t$ is zero~\cite{demler}. The close 
similarity between the phenomenological effective theory of a phase-coherent bilayer system and that 
of a Josephson junction has led to suggestions that the phase-coherent bilayers should exhibit a 
dc Josephson effect in interlayer tunneling measurements~\cite{wenzee,izawa1,izawa2,fogler}.

The first {\it direct} experimental signature of the phase-coherent state and the linearly 
dispersing collective mode was observed by Spielman {\it et al.} who discovered a dramatic 
enhancement 
of the zero-bias tunneling conductance $G_T$ as the layer separation $d$ (measured in units of 
magnetic length) was reduced~\cite{spielman1,spielman2}. Although this dramatic enhancement is 
reminiscent of the tunneling characteristics of a Josephson junction, it is not yet clear 
whether the finite width and height of the peak is intrinsic, or is governed by the temperature and 
experimental limitations. 

Here we present a theory of the zero-bias conductance $G_T$ that is non-perturbative in the 
tunneling amplitude $\Delta_t$. We find that the tunnel conductance $G_T$ is given by Fermi's 
Golden rule for the {\it quasiparticles} in the phase-coherent state, and that the breakdown of 
Fermi's Golden rule for the bare electrons is therefore expected. In the following sections we 
first discuss the physical picture that underlies our theory, and then present the results of an 
approximate but fully microscopic calculation. The details of this calculation are available 
elsewhere~\cite{ynjprl}.

\noindent{\it Physical Picture:} Let us consider a bilayer quantum Hall system at total filling 
factor $\nu=1$ with tunneling 
amplitude $\Delta_t$. In the pseudospin picture, the tunneling amplitude acts as a field along 
the $x$-axis in the pseudospin-space, whereas a bias-voltage or a random disorder potential acts 
as a field along the $z$-axis. In the mean-field approximation, the interlayer exchange Coulomb 
interaction enhances the splitting between symmetric and antisymmetric single-particle energies, 
$\Delta_{qp} = \Delta_t + M_0\Delta_{sb}$, where $M_0$ is the dimensionless order-parameter for 
the phase-coherent state. This exchange-enhancement $M_0\Delta_{sb}$ survives in the limit 
$\Delta_t\rightarrow 0$ and gives rise to spontaneous interlayer phase-coherence~\cite{demler}. 
We will call the exchange-enhanced splitting $\Delta_{qp}$ as the quasiparticle tunneling 
amplitude.

The tunneling amplitude $\Delta_t$ is the only term in the microscopic Hamiltonian that does not 
conserve the charge in each layer separately. Recent theoretical 
work~\cite{fogler,balents,stern,ynjsces} 
has shown that the bare-electron Golden rule used to estimate the zero-bias conductance $G_T$ breaks 
down 
as $\Delta_t \to 0$, {\it i.e.} that $\lim_{\Delta_t \to 0} G_T/\Delta_t^2$ diverges in the 
phase-coherent state. Evaluation of  $G_T$ therefore requires a non-perturbative theory like the one 
we present here. Our finding, that $G_T$ is finite once the presence of low-energy quasiparticle 
excitations at finite temperature $T\ne 0$ and with disorder even at $T=0$ is acknowledged, is at 
odds with the prediction that this system should show a dc Josephson effect. In usual dc Josephson 
effect systems, the persistent currents are maintained by a current-direction order-parameter 
phase-slip barrier that has no analog in bilayer quantum Hall systems.

Now let us examine the possibility of persistent currents in a bilayer system when we treat the 
interlayer tunneling non-perturbatively. First we recall that the interlayer current operator is 
given by $\hat{I} = eNM_0\Delta_t \hat{M}_y/\hbar$, where $N$ is the total number of electrons. 
Thus, a current-carrying initial state corresponds to the pseudospin lying in the $x$-$y$ plane 
with a nonzero $y$-component and a concurrent field $\Delta_t$ along the $x$-axis. Since the 
bilayer system has an easy-plane anisotropy, it is clear that in the absence of a bias-voltage the 
pseudospin 
must relax and point along the $x$-axis or, equivalently, that the initial-state current must 
decay. In other words, in the absence of bias-voltage, a state with nonzero interlayer current 
cannot be a steady state. {\it This physical picture underlies our theory, and provides a 
transparent way to see why persistent currents are impossible in the presence of a finite 
interlayer tunneling.}

In a steady state the rate of change of interlayer current is zero. This constraint allows us to 
express the zero-bias tunnel conductance $G_T$ in terms of a phenomenological lifetime $\tau$ 
for the interlayer current,
\begin{equation}
\label{eq: one}
G_T = \frac{e^2NM_0\Delta_t\tau}{\hbar^2}.
\end{equation}
At a microscopic level, there are several possible channels for the decay of the current; 
disorder broadening of the quasiparticle bands, local density fluctuations leading to puddles of 
compressible regions, {\it etc}. In the present calculation, we assume that the broadening of the 
mean-field quasiparticle bands because of disorder leads to a finite density of states at the 
Fermi energy and provides a channel for the current to decay into particle-hole pairs. It is 
then possible to extract the relaxation time $\tau$ for the interlayer current from the 
dynamical response function $\chi_{yz}(\omega)$,
\begin{equation}
\label{eq: two}
\tau = M_0 \Re[\chi_{yz}^{-1}](\omega\rightarrow 0),
\end{equation}
and obtain the dependence of the zero-bias conductance $G_T$ on interlayer tunneling amplitude 
$\Delta_t$ by using Eq.(\ref{eq: one}).

\noindent{\it Results and Discussion:} We evaluate the response function $\chi_{yz}$ using 
disorder vertex corrections and the 
generalized random phase approximation, which capture the physics of collective excitations in 
the presence of a random disorder potential. A straightforward calculation~\cite{ynjprl} shows 
that the zero-bias tunnel conductance is given by the Fermi's Golden rule for the mean-field 
quasiparticles
\begin{equation}
\label{eq: three}
G_T \propto \Delta_{qp}^2 = (\Delta_t + M_0\Delta_{sb})^2.
\end{equation}

\begin{figure}[h]
\begin{center}
\hspace{0.2cm}
\epsfxsize=3.5in
\epsffile{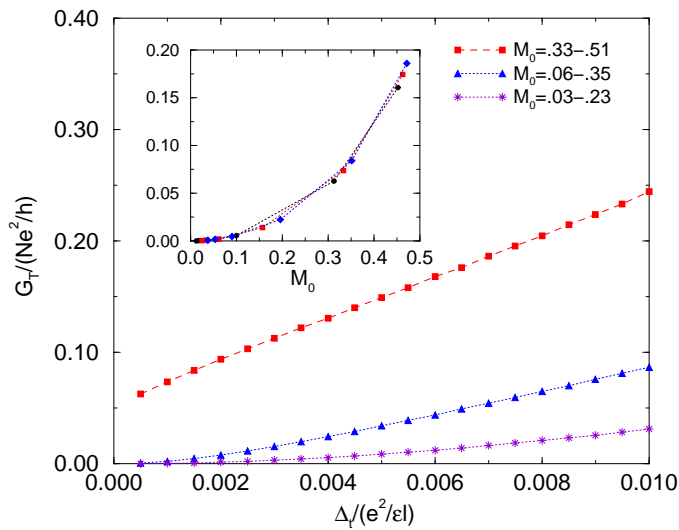}
\caption{Zero-bias conductance $G_T$ as a function of interlayer tunneling amplitude $\Delta_t$ 
for various disorder strengths. The strength of disorder is characterized by the suppression of 
the order parameter $M_0$ from its clean limit value $M_0=1$. The inset shows the $\Delta_t\rightarrow 0$ limit of $G_T$ as a function of the order parameter $M_0$.} 
\label{fig: conductance}
\end{center}
\end{figure}

Figure~\ref{fig: conductance} shows the results for the zero-bias conductance $G_T$ as a 
function of interlayer tunneling amplitude $\Delta_t$ for various disorder strengths. We 
characterize the 
strength of disorder by the suppression of the mean-field order parameter $M_0$ from its clean 
limit value, $M_0 =1$. When the system is barely phase-coherent ($M_0<<1$) we find that 
$G_T \propto \Delta_t^2$; however, as the interlayer phase-coherence develops, the conductance 
$G_T$ remains finite as $\Delta_t\rightarrow 0$. This result is clearly non-perturbative in the 
tunneling amplitude $\Delta_t$, and consistent with the breakdown of Fermi's Golden rule for the 
bare electrons. 

The inset in Fig.~\ref{fig: conductance} shows the dependence of the zero-tunneling-amplitude 
conductance $G_T(\Delta_t\rightarrow 0)$ on the order-parameter $M_0$. We see that the quadratic 
dependence is consistent with the general result, Eq.(~\ref{eq: three}).  

Thus, we find that the zero-bias conductance of a bilayer system is finite and proportional to the 
square of the quasiparticle tunneling amplitude. This result provides further support for the idea that 
a weak-coupling Hartree-Fock description of the phase-coherent state is indeed reliable.

\noindent{\it Acknowledgment:} This work was supported by the Welch Foundation, the Indiana 21st 
Century Fund, and the NSF grant no. DMR 0115947.

%===============================================================================================%

%===============================================================================================%

\end{document}